\documentclass[aps,twocolumn,reprint,amsmath,amssymb,showpacs,superscriptaddress]{revtex4-1}
\usepackage{multirow}
\usepackage{graphicx,epsfig}
\usepackage{epstopdf}
\usepackage{wasysym}
\usepackage{stackengine}
\usepackage{color,soul} 
\usepackage[colorlinks=true,citecolor=blue,linkcolor=red,linktocpage=true,pagebackref=false]{hyperref}

\begin{document}
\newcommand{\be}{\begin{equation}}
\newcommand{\ee}{\end{equation}}
\newcommand{\ba}{\begin{eqnarray}}
\newcommand{\ea}{  \end{eqnarray}}
\newcommand{\ve}{\varepsilon}
\newcommand{\Tau}{\mathcal{T}}
\newcommand{\calb}{\mathcal{B}}
\newcommand{\calz}{\mathcal{Z}}
\newcommand{\calp}{\mathcal{P}}
\newcommand{\cale}{\mathcal{E}}
\newcommand{\cald}{\mathcal{D}}
\title{Enhanced performance of a quantum-dot-based nanomotor due to  Coulomb interactions}
%\title{Interaction-enhanced performance of adiabatically driven quantum dots operating as quantum motors}
%MC Alternative title commented
\author{Mar\'{\i}a Florencia Ludovico}
\affiliation{Consiglio Nazionale delle Ricerche, Istituto Officina dei Materiali (IOM) and Scuola Internazionale Superiore di Studi Avanzati (SISSA),
Via Bonomea 265, I-34136, Trieste, Italy}
\author{Massimo Capone}
\affiliation{Consiglio Nazionale delle Ricerche, Istituto Officina dei Materiali (IOM) and Scuola Internazionale Superiore di Studi Avanzati (SISSA),
Via Bonomea 265, I-34136, Trieste, Italy}

\begin{abstract}
We study the relation between quantum pumping of charge and the work exchanged with the driving potentials in a strongly interacting {\it{ac}}-driven quantum dot. We work in the large-interaction limit and in the adiabatic pumping regime, and we develop a treatment that combines the time-dependent slave-boson approximation with linear response in the rate of change of the {\it{ac}}-potentials. We find that the time evolution of the system can be described in terms of equilibrium solutions at every time. We analyze the effect of the electronic interactions on the performance of the dot when operating as a quantum motor. 

The main two effects of the interactions are a shift of the resonance and an enhancement of the efficiency with respect to a non-interacting dot. This is due to the appearance of additional {\it{ac}}-parameters accounting for the interactions that increase the pumping of particles while decreasing the conductance.

\end{abstract}

% 73.23.-b Electronic transport in mesoscopic systems 
% 72.10.Bg General formulation of transport theory 
% 73.63.Kv Quantum dots 
% 44.10.+i Heat conduction

\maketitle

\section{Introduction} 

The understanding and description of energy conversion processes in nanoscale devices is crucial for the development of novel nanotechnologies. In particular, the manipulation of charge and energy fluxes, and the control of energy dissipation are strategic tasks for the design of energy-efficient circuits. Nanostructures working at low temperatures, as for example, quantum dots, are perfect candidates because the energy-filter effect is maximized by the discrete energy spectrum and their electronic and optical properties can be controlled through suitable changes of composition, size and shape \cite{benenti,Dwyer}. 

Previous studies have shown remarkable performances of quantum dots in thermoelectric devices \cite{beenakker, Rafael,jordan,karlstrom}, in which the conversion between electrical and thermal energy takes place due to the application of {\it{dc}}-drivings through both temperature and voltage biases. The thermoelectric properties and performance as heat-engines of quantum dots systems have been widely addressed \cite{david,kuo,thierschmann,roura3,roura,cosi,liu,rejec,muralidharan,rosadavid,taylor,andreev,sierra,erdman,rossello,liu2,sierra2,weymann,boese,scheibner,karki,roura2,dutt}. The small size of these setups makes quantum interference and Coulomb interaction important. The effect of the latter has been considered \cite{cosi,liu,rejec,muralidharan,rosadavid,taylor} either in the Coulomb-blockade \cite{andreev,sierra,erdman,rossello,liu2,sierra2} or Kondo regimes \cite{weymann,boese,scheibner,karki,roura2}, mostly within the linear response regime and, to a lesser extent, for heat engines operating in the nonlinear regime \cite{rosadavid,sierra,erdman,karki,roura2,dutt}. It has been indeed reported \cite{erdman,liu2,scheibner,karki} that the presence of Coulomb interactions leads to an improvement of the thermoelectric performance in quantum dots by enhancing the thermopower and decreasing the thermal conductance. 

Another route to boost the thermoelectric response is the application of time-dependent gate voltages.
Refs. \cite{adeline,hanggi} report an enhancement of the thermopower during the transient regime following a sudden change of the gating potential.
The thermoelectric response under {\it{ac}}-drivings (time-periodic) has been studied in \cite{ac1,ac2,lim}.

All these studies have focused on quantum-dot-based devices converting heat into electricity, or vice versa.  Nevertheless, when {\it{dc}} and {\it{ac}} drivings are applied at the same time, an exchange of energy between the different kinds of driving sources can occur. In this way, the system not only allows for thermoelectric effects, but it can also behave as a quantum machine that transforms electrical or thermal energy into another form of energy, that could be for example mechanical work \cite{bustos}. Consequently, regarding this latter kind of energy conversion, other operational modes appear: (i) {\it{Quantum motors}} and {\it{generators}}, when the system is driven by a bias voltage together with the application of {\it{ac}}-potentials; (ii) {\it{Heat engines}} and {\it{heat pumps}}, when the device involves a temperature gradient instead of a bias voltage.

The response of nanostructures working in these last operational modes has been less studied in the literature, and therefore, the search for mechanisms boosting their efficiency is still an open field.  Among the first works, Ref. \cite{juergens} considered a driven double quantum dot operating as a generator or a heat pump/heat engine, while in  Refs. \cite{bustos,bruch} quantum motors based on Thouless pumps have been discussed. Furthermore, a cold-atom-based {\it{ac}}-driven quantum motor was explored in Ref. \cite{hanggi2}.

The performance of such nanomotors and nanoengines is based on the quantum pumping effect \cite{thouless,brouwen}, that consists in the generation of a {\it{dc}} current at zero bias voltage, by merely applying local {\it{ac}}-drivings to a quantum coherent conductor. When the driving period is much larger than any other characteristic time scale of
the system, the pumping is called {\it{adiabatic}}. 
The key for the operation of these machines is built on the relation between the output power $P_{out}$ and the charge (for a motor/generator) or heat (for heat engines/pumps) pumped through the system during one period of the {\it{ac}}-driving. In the case of an adiabatic quantum motor it reads \cite{bustos}
\begin{equation}\label{power}
P_{out}=Q_p\,V/\tau,
\end{equation}
where $Q_p$ is the pumped charge per period, $V$ is the applied bias voltage and $\tau$ is the period of the {\it{ac}}-driving. The above equation lays out the fact that nanomotors are realized by the simultaneous application of {\it{dc}} and {\it{ac}} drivings, since no output power is obtained either at zero bias voltage or without pumping of charge. 
On the other hand, an extension of the linear response theory of thermoelectrics to systems under adiabatic {\it{ac}}-driving has been recently presented in Ref. \cite{ludovico}. This theory describes the relation between heat and particle fluxes, and the energy flux exchanged between the electronic system and the {\it{ac}}-driving sources, through Onsager reciprocity relations. 
This theory also allows for the characterization of all the operational modes in (i) and (ii) in terms of efficiencies and figures of merit.

\begin{figure}[t]
  \includegraphics[width=0.4\textwidth]{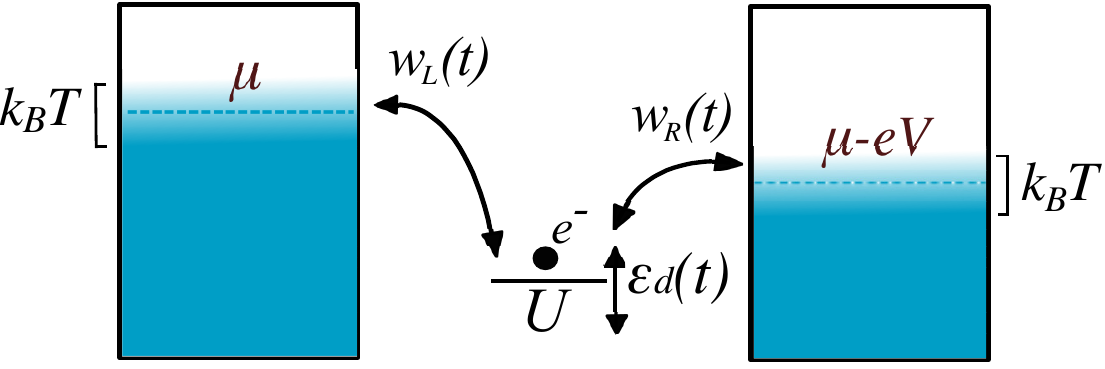}
  \caption{Scheme of the device. We take a single-level quantum dot connected to two non-interacting electronic reservoirs at the same temperature $T<T_{Kondo}$. A constant bias voltage $V$ is applied between the reservoirs, so that the chemical potential is $\mu_L=\mu-eV$ on the left  and $\mu_R=\mu$ on the right. The hopping elements between the dot and the reservoirs, $w_L(t)$ and $w_R(t)$, and the energy level of the dot $\ve_d(t)$ are time-periodic functions. %In this work, we are going to consider the quantum dot in the strongly interacting limit $U\rightarrow \infty$. 
  }\label{fig1}
\end{figure}

The goal of the present work is to study the effects of the Coulomb interactions on the performance of an adiabatically {\it{ac}}-driven quantum dot operated as a {\it{motor}} within the framework of Ref. \cite{ludovico}. Fig. \ref{fig1} shows the setup we have in mind. The quantum dot is coupled to two non-interacting reservoirs at the same temperature, but with different chemical potentials. The time-periodic driving is introduced through the tunneling elements $w_\alpha (t)$, with $\alpha=L,R$, as well as  by a modulation of the energy level of the dot $\ve_d(t)$.
The behavior of the device as a {\it{motor}} can be understood as follows: when the {\it{ac}}-driving pumps particles into the reservoir with lower chemical potential, the gain in electrical energy can be used to perform work on the {\it{ac}}-sources. This work can be later transformed in, e.g., mechanical energy. 
%%%%%%%%%%
 %%%%%%%%%%

In order to describe the interacting quantum dot within the adiabatic regime, we develop a treatment that combines the time-dependent slave-boson mean-field approximation \cite{wu,citro,citro2} with a linear-response expansion in the small rate of change of the {\it{ac}}-parameters \cite{ludovico}. Within our approach we can  describe the interacting dot at every instant of time in terms of frozen {\it{equilibrium}} solutions, as if we were taking a sequence of snapshots of the system. We note that this technique can be extended to a more generic setup and also beyond linear response in the adiabatic approximation. 

The paper is organized as follows. In Sec. \ref{sec2} we introduce the model and the time-dependent slave-boson mean-field. Sec. \ref{formalism} presents the solution of the slave-boson approximation within the adiabatic response regime. In Sec. \ref{nanomotors} we apply this formalism to study the performance of {\it{ac}}-driven quantum dots as nanomotors. 
In Sec. \ref{Results}, taking the case $\ve_d(t)=0$ as an illustrative example, we analyze in detail the effects of the electron-electron interactions on the linear-response transport coefficients and on the efficiency at maximum power. Finally Section \ref{conclusions} is devoted to the summary and conclusions.

\section{Model and formalism}\label{sec2}

We consider a simple setup featuring all the necessary ingredients to analyze the effects of the electron-electron interaction in quantum dots-based motors. It consists in a single level quantum dot connected to two non-interacting electronic reservoirs at the same finite temperature $T$. The setup is shown in Fig. \ref{fig1}. Charge and energy fluxes in the system are driven by both an applied small bias voltage $V$ between the leads, and the time-periodic modulation of the energy level $\ve_d(t)$ and the couplings $w_\alpha (t)$. The time-dependent driving can be implemented by the local application of three {\it{ac}}-gate voltages, one for shifting the energy of the level, while the other two for controlling the transparency of the tunneling barriers. We allow only the module of $w_\alpha$ to vary in time, but not its phase, since a time-dependent phase would correspond to a time-dependent bias voltage. We assume an adiabatic {\it{ac}}-driving, which corresponds to a driving period much larger than the typical dwell time for the electrons inside the driven structure. Being $\omega$ the typical frequency of the periodic drivings, and $\Gamma$ the hybridization with the reservoirs, the adiabatic condition requires $\hbar\omega\ll \Gamma$ \cite{brouwen}. 

The system is described by the Hamiltonian 
\be\label{hamtot}
H(t)= H_d(t)+H_\mathcal{T}(t)+H_{leads},
\ee
where the first term corresponds to the interacting quantum dot 
\be\label{hamdot}
H_d(t)=\sum_\sigma \ve_d(t)\, d^\dagger_\sigma d_\sigma 
+ U n_{\uparrow}n_{\downarrow},
\ee
with $d^\dagger_\sigma$ and $d_\sigma$ being, respectively, the creation and destruction operators of an electron with spin $\sigma=\uparrow, \downarrow$. The second term, that depends on the occupation number operator $n_\sigma=d^\dagger_\sigma d_\sigma$ describes the Coulomb repulsion between electrons, with an interaction energy $U$. The non-interacting leads are represented by the  Hamiltonian 
\be\label{hamres}
H_{leads}=\sum_{\alpha=L, R}\sum_{k_\alpha,\sigma}\ve_{k_\alpha} c^\dagger_{k_\alpha\sigma} c_{k_\alpha\sigma},
\ee
where $\ve_{k_\alpha}$ is the energy band of the reservoir $\alpha$ and the operator $c^\dagger_{k_\alpha\sigma}$ ($c_{k_\alpha\sigma}$) creates (destroys) an electron with momentum $k_\alpha$ and spin $\sigma$ in the reservoir $\alpha$. Finally, the coupling between the dot and the reservoirs is represented by the following tunneling Hamiltonian
\be\label{hamtunneling}
H_{\mathcal{T}}(t)=\sum_{\alpha,k_\alpha,\sigma}w_{\alpha}(t) c^\dagger_{k_\alpha\sigma}d_\sigma +h.c.
\ee

\subsection{Time-dependent slave-boson approximation}

In this work we focus on the strongly interacting limit $U\rightarrow \infty$, which we address by means of the time-dependent slave-boson mean-field theory \cite{wu,citro,citro2}, which extends Coleman slave-boson equilibrium approach \cite{coleman} to the case of {\it{ac}}-driven quantum dots. 
We choose the slave-boson mean-field approach as a minimal theoretical framework which captures the main effect of strong correlations but still allows for an analytical treatment which is easily combinable with our linear-response framework. The present studied can be seen as the basic brick for more advanced treatments employing accurate solutions of the quantum dot problem. An obvious extension would be the use of the Kotliar-Ruckenstein formalism to consider the system at finite values of $U$ \cite{KR}.

Within the slave boson formalism the fermion in the strongly interacting limit can be represented in terms of a bosonic field $b$ and a pseudofermionic operators $f_\sigma$. Then, the operators of the quantum dot can be written in terms of the bosonic and quasi-fermionic operators as: $d_\sigma\rightarrow b^\dagger f_\sigma$, $d^\dagger_\sigma\rightarrow f^\dagger_\sigma b$. Plugging these expressions into Eq. (\ref{hamtot}), the slave-boson Hamiltonian of our system can be written as
\ba
H_{SB}(t)&=& H_{leads}+\sum_\sigma (\ve_d(t)+\lambda(t))f_\sigma^\dagger f_\sigma\nonumber\\
&&+\sum_{\alpha,k_\alpha,\sigma}w_{\alpha}(t)\,b^\dagger c^\dagger_{k_\alpha\sigma}f_\sigma +h.c\\
&&+\lambda(t)(b^\dagger b -1)\nonumber,
\ea
where $\lambda(t)$ is a Lagrange multiplier enforcing the constraint preventing double occupancy at any time
\be\label{constUinfty}
N_d(t) +\langle b^\dagger b\rangle -1=0,
\ee
where  ${N}_d=\sum_\sigma\langle f^\dagger_\sigma f_\sigma\rangle$. For our problem the Lagrange multiplier $\lambda$ will be time-periodic due to the application of the external {\it{ac}}-potentials. The bosonic operator $b$ evolves according to the equation of motion
\be
i\hbar\partial_t b=[b, H_{SB}]=\lambda(t) b +\sum_{\alpha,k_\alpha,\sigma} w_{\alpha}(t)\,c^\dagger_{k_\alpha\sigma}f_\sigma.
\ee

As customary, we treat the slave-boson operator in the mean-field (MF) approximation replacing the salve-boson operator $b$ with its expectation value $\langle b\rangle=\calb(t)$, while neglecting its fluctuations. This assumption is justified within the adiabatic regime, in which the evolution of the system is quasi-static, so that the slow variation of the {\it{ac}}-potentials do not significantly affect the condensation of the bosons. As a consequence of the MF approximation, the original problem turns out to be described by a constrained {\it{non-interacting}} theory with the time-dependent MF Hamiltonian 
\ba\label{hammf}
H_{SB}^{MF}(t)&=& H_{leads}+\sum_\sigma (\ve_d(t)+\lambda(t))f_\sigma^\dagger f_\sigma\nonumber\\
&&+\sum_{\alpha,k_\alpha,\sigma}w_{\alpha}(t)\,\calb^*(t) c^\dagger_{k_\alpha\sigma}f_\sigma +h.c\\
&&+\lambda(t)(\vert \calb(t)\vert^2 -1)\nonumber,
\ea  
and the following set of non-linear equations to be solved in order to find the unknown parameters $\lambda(t)$ and $\calb(t)$
\begin{subequations}\label{constraints}
\ba
\lambda(t)\calb(t) +\sum_{k_\alpha,\alpha,\sigma}w_{\alpha}(t)\langle c^\dagger_{k_\alpha\sigma}f_\sigma\rangle &= & i\hbar\partial_t\calb (t)\label{a}\\
N_d(t) +\vert\calb(t)\vert^2-1 & = &
0.\label{num}
\ea
\end{subequations}

\section{Dynamics within the adiabatic response regime}\label{formalism}

We now develop a method to solve the non-linear set of equations in (\ref{constraints}) within the adiabatic approximation. In this regime, all the {\it{ac}}-potentials slowly evolve in time, being their rates of change %$\dot{\ve}_d$ and $\dot{w}_\alpha$
very small (while their amplitude can be arbitrary) which allows us to keep only the contributions up to first order in the temporal variation of the {\it{ac}}-drivings. 
Within this linear-response approximation, the slave-boson field $\calb (t)$ can be taken as a real number, as it is generally considered in the stationary case. Nevertheless, it is important to mention that the phase must be considered to work beyond linear response in the rate of change of the {\it{ac}}-potentials. We can start by multiplying the equation of motion in (\ref{a}) by $\calb(t)$, and separating it into its real part  
\be\label{realpart}
\lambda(t)\calb^2(t) +\sum_{\alpha,k_\alpha,\sigma}\mbox{Re}\left\{ w_{\alpha}(t)\calb(t)\langle c^\dagger_{k_\alpha\sigma}f_\sigma\rangle\right\} =0,
\ee
and imaginary part
\be\label{imaginarypart}
\frac{1}{\hbar}\sum_{\alpha,k_\alpha,\sigma}2\,\mbox{Im}\left\{w_{\alpha}(t)\calb(t)\langle c^\dagger_{k_\alpha\sigma}f_\sigma\rangle\right\}=\partial_t\calb^2(t).
\ee
This leaves us with three equations to be solved, (\ref{num}), (\ref{realpart}) and (\ref{imaginarypart}), and two unknown parameters. This means that the system of equations is overdetermined. In fact, we find that Eqs. (\ref{imaginarypart}) and (\ref{num}) are related. In order to show that, we notice that the left-hand side of the above equation is exactly the total flux of particles entering the leads $\dot{N}_{leads}(t)=\sum_{\alpha=L,R}\dot{N}_\alpha(t)$, with
\ba\label{particlecurrent}
\dot{N}_{\alpha}(t)&=&\sum_{k_\alpha,\sigma} \frac{i}{\hbar}\langle[H_{SB}^{MF},c_{k_\alpha\sigma}^\dagger c_{k_\alpha\sigma}]\rangle\\
&& = \frac{1}{\hbar}\sum_{k_\alpha,\sigma}2\,\mbox{Im}\left\{w_{\alpha}(t)\calb(t)\langle c^\dagger_{k_\alpha\sigma}f_\sigma\rangle\right\}.\nonumber
\ea
Due to the conservation of the number of particles of the full system, the current entering the leads must be equal to the one leaving the dot, $\dot{N}_{leads}(t)=-\dot{N}_{d}(t)$. Therefore Eq.(\ref{imaginarypart}) can be written as $\dot{N}_{d}(t)+\partial_t\calb^2(t)=0$, that is just the derivative in time of the constraint on the occupation of the dot in Eq. (\ref{num}). This means that within the adiabatic response regime, the set of equation in (\ref{constraints}), can be reduced to a $2\times 2$ system containing only Eqs. (\ref{num}) and (\ref{realpart}).

In order to express this system in terms of the variables $\lambda(t)$ and $\calb(t)$, we can follow the steps detailed in Appendix \ref{appendix1} for slow adiabatic drivings, and write the expectation value $\langle c^\dagger_{k_\alpha\sigma}f_\sigma\rangle(t,t)$ in terms of the Fourier transforms of the retarded Green function $G^r_\sigma(t,t')=-i\theta(t-t')\langle\{f_\sigma(t),f^\dagger_\sigma(t')\}\rangle$ and lesser Green function $G^<_\sigma(t,t')=i\langle f^\dagger_\sigma(t')f_\sigma(t)\rangle$ of the quantum dot
\be\label{greensdot}
G^{r,<}_\sigma(t,t')=\int^{\infty}_{-\infty}\frac{d\ve}{2\pi}e^{-i\frac{\ve}{\hbar}(t-t')}G^{r,<}_\sigma(t,\ve),
\ee
which can be obtained by solving a Dyson equation \cite{dyson}. The lesser Green function is related to the occupation number of the dot through $N_d(t)=-i\sum_\sigma G^<_\sigma(t,t)$ .

In this way the slave-boson equations read
\begin{subequations}\label{greenset}
\ba
\lambda(t)\calb^2(t) & = & \!-\!\!\sum_{\alpha}\!\int\!\frac{d\ve}{\pi}\!\left[\tilde{\Gamma}_\alpha(t) \mbox{Re}\!\left\{\!G^r(t, \ve)f_\alpha(\ve)+\frac{G^<(t,\ve)}{2}\!\!\right\}\right.\nonumber\\
 && \left.-\frac{\hbar}{2}\,\dot{\tilde{\Gamma}}_\alpha(t) \,\mbox{Im}\left\{\partial_\ve G^r(t,\ve)\right\}f_\alpha(\ve)\right]\label{greenset1}\\
\calb^2(t)-1&=&-\int\frac{d\ve}{\pi}\mbox{Im}\left\{G^<(t,\ve)\right\},
\ea
\end{subequations}
where we have assumed a spin-symmetric solution $G^{r,<}\equiv G^{r,<}_\uparrow=G^{r,<}_\downarrow$. The function $f_\alpha(\ve)=[e^{(\ve-\mu_\alpha)/(k_BT)}+1]^{-1}$ is the Fermi-Dirac distribution of the reservoir $\alpha$, with $k_B$ being the Boltzmann's constant.  We have also defined $\tilde{\Gamma}_\alpha(t)=\calb^2(t)\Gamma_\alpha(t)$ as the renormalized hybridization of the lead $\alpha$ due to the interactions, where $\Gamma_\alpha(t)=\vert w_\alpha(t)\vert^2\rho_\alpha$ is the hybridization in the non-interacting case and $\rho_\alpha$ is the density of states of the lead. We consider the wide-band limit, in which the densities $\rho_\alpha$, and therefore the hybridizations ${\Gamma}_\alpha$, are energy-independent. It is important to notice that the derivative $\dot{\tilde{\Gamma}}_\alpha$ involves not only the small rate of change of tunneling elements, inside $\dot{\Gamma}_\alpha$, but also the temporal variation of the bosonic field $\dot{\calb}$. 
As a result of the slow evolution in time of the applied {\it{ac}}-drivings, the temporal variation of the parameters accounting for the interactions, $\dot{\lambda}$ and $\dot{\calb}$, are also small. For this reason, in Eq. (\ref{greenset1}), we only keep the terms up to linear order in $\dot{\tilde{\Gamma}}_\alpha$. 

We can obtain exact results by expanding $G^r(t,\ve)$ and $G^<(t,\ve)$ up to first order in the temporal variation of all the {\it{ac}}-parameters of the MF Hamiltonian (\ref{hammf}) \cite{lowfreq1,lowfreq2} (see Appendix \ref{lowfreq} for details). Moreover, we can also evaluate the integrals of Eq. (\ref{greenset}) in linear response in the small bias voltage by expanding the Fermi distribution as $f_R\sim f+eV\partial_\ve f$, with $f=f_L$. Then, 
\begin{subequations}\label{finalsetrho}
 \ba\label{finallambda}
 \lambda(t)\calb^2(t) & = & - \int\frac{d\ve}{\pi}f(\ve)\rho(t, \ve)(\ve-\tilde{\ve}_d(t))\nonumber\\
 &&-\int\frac{d\ve}{\pi}\partial_\ve f(\ve)\rho(t, \ve)(\ve-\tilde{\ve}_d(t))\times\\
 &&\left[eV\frac{{{\Gamma}}_R(t)}{{\Gamma}(t)}+\frac{\hbar}{2}\rho(t, \ve)\tilde{\Gamma}(t)\partial_t\!\!\left(\frac{(\ve-\tilde{\ve}_d(t))}{\tilde{\Gamma}(t)}\right)\right],\nonumber
 \ea
 and 
 \ba\label{finaln}
\calb^2(t)-1 & = & -\int\frac{d\ve}{\pi}f(\ve)\rho(t, \ve)\nonumber\\
 &&-\int\frac{d\ve}{\pi}\partial_\ve f(\ve)\rho(t, \ve)\times\\
 &&\left[eV\frac{{\Gamma}_R(t)}{{\Gamma}(t)}+\frac{\hbar}{2}\rho(t, \ve)\tilde{\Gamma}(t)\partial_t\!\left(\frac{(\ve-\tilde{\ve}_d(t))}{\tilde{\Gamma}(t)}\right)\right],\nonumber
 \ea
 \end{subequations}
where $\tilde{\Gamma}(t)=\calb^2(t)\Gamma(t)$, with $\Gamma(t)=\Gamma_L(t)+\Gamma_R(t)$, is the total hybridization, $\tilde{\ve}_d(t)=\ve_d(t)+\lambda(t)$ is the renormalized energy level of the quantum dot, and $\rho(t,\ve)=\tilde{\Gamma}(t)[(\ve-\tilde{\ve}_d(t))^2+(\tilde{\Gamma}(t)/2)^2]^{-1}$ corresponds to the density of states describing the regime in which the electrons instantaneously adjust its potential to the {\it{ac}}-fields. The first terms on the right hand side in Eqs. (\ref{finallambda}) and (\ref{finaln}) describe the system as being at equilibrium \cite{dong} at every {\it{frozen}} time $t$, while the last ones represent the corrections due to the small {\it{dc}}-driving $V$ and the slow (but not necessarily weak or small) {\it{ac}}-drivings.

\subsection{Leading-order solutions}
For low driving frequencies $\omega$ and small bias voltages $V$, we propose the following solution for the non-linear system of equations in Eq. (\ref{finalsetrho}):
\be\label{sollambda}
\lambda(t)=\lambda^f(t)+\Delta\lambda^{V}(t)+\Delta\lambda^\omega(t),
\ee
and
\be\label{solb2}
\calb^2(t)={\calb^f}^2(t)+{\Delta\calb^2}^V(t)+{\Delta\calb^2}^\omega(t),
\ee
since Eqs. (\ref{finallambda}) and (\ref{finaln}) depend quadratically on the  bosonic expectation value. 
Here, $\lambda^f(t)$ and ${\calb^f}^2(t)$ are the {\it{frozen}} solutions considering that the system is at equilibrium at every time, as in a sequence of snapshots. They are obtained by solving numerically the following non-linear system of equations
\begin{subequations}\label{setfrozen}
\be
\lambda^f(t)\calb^{f^2}(t)=- \int\frac{d\ve}{\pi}f(\ve)\rho^{f}(t, \ve)(\ve-\tilde{\ve}_d^f(t))
\ee
\be
\calb^{f^2}(t)-1=- \int\frac{d\ve}{\pi}f(\ve)\rho^{f}(t, \ve),
\ee 
\end{subequations}
where $\rho^f=\rho(\lambda=\lambda^f,\calb^2=\calb^{f^2})$ is the {\it{frozen}} density of states, and $\tilde{\ve}_d^f=\ve_d+\lambda^f$.  $\Delta\lambda^{V,\omega}$ and ${\Delta\calb^2}^{V,\omega}$ are the corrections due to the presence of the bias voltage $V$ and the {\it{ac}}-drivings with frequency $\omega$. Since the time derivative of the {\it{ac}}-potentials is proportional to the driving frequency, the first-order corrections are $\Delta\lambda^{\omega},\,{\Delta\calb^2}^{\omega}\propto \omega$. Similarly, $\Delta\lambda^{V},\,{\Delta\calb^2}^{V}\propto V$.

To compute the corrections we have to plug Eqs. (\ref{sollambda}) and (\ref{solb2}) into (\ref{finalsetrho}), and perform an expansion up to first order in $\Delta\lambda^{V,\omega}$ and ${\Delta\calb^2}^{V,\omega}$. After that, we find two independent $2\times 2$ systems of linear equations of the form
\be\label{systemcorrections}
\sum_{j=1}^2 M_{ij}(t)X_j^{\beta}(t)=C_i^\beta(t)\,\,\,\,\text{for}\,\,\,\,i=1,2\
\ee
where $\beta=V, \omega$, as before, is an index indicating the nature of the correction, i.e. if it is due to the {\it{dc}}-driving or the {\it{ac}}-drivings. The vectors $\vec{X}^\beta$ collect the unknown corrections of both the Lagrange multiplier and the bosonic field,
$\vec{X}^{V}=(\Delta\lambda^{V},{\Delta\calb^2}^{V})$ and $\vec{X}^{\omega}=(\Delta\lambda^{\omega},{\Delta\calb^2}^{\omega})$. The matrix $\hat{M}$ contains the coefficients of the system
\ba\label{m}
M_{11}(t)&=&M_{22}(t)=1+\!\int\frac{d\ve}{\pi}\partial_\ve f \frac{\rho^f(t,\ve)}{{\calb^f}^2(t)}(\ve-\tilde{\ve}_d^f(t))\nonumber\\
M_{21}(t)&=&-M_{12}(t)\left(\frac{2}{\Gamma(t)}\right)^2=\int\frac{d\ve}{\pi}\partial_\ve f\,\rho^f(t,\ve),
\ea
and, finally, the components $\vec{C}_i^{V,\omega}$ with $i=1,2$, are the independent terms 
\ba\label{c}
C_i^{V}(t)&=&-eV\frac{{\Gamma}_R(t)}{{{\Gamma}}(t)}\int\frac{d\ve}{\pi}\partial_\ve f\rho^f(t, \ve)\left(\frac{(\ve-\tilde{\ve}^f_d(t))}{{\calb^f}^2(t)}\right)^{(2-i)}\nonumber
\\
C_i^{\omega}(t)&=& -\frac{\hbar}{2}\tilde{\Gamma}^f(t)\int\frac{d\ve}{\pi}\partial_\ve f{\rho^f}^2(t, \ve)\times\\
&&\partial_t\!\!\left(\frac{(\ve-\tilde{\ve}^f_d(t))}{\tilde{\Gamma}^f(t)}\right)\left(\frac{(\ve-\tilde{\ve}^f_d(t))}{{\calb^f}^2(t)}\right)^{(2-i)},\nonumber
\ea
where $\tilde{\Gamma}^f={\calb^f}^2\Gamma$.
The solutions of the systems of equation in (\ref{systemcorrections}) read
\be\label{correct1}
\Delta\lambda^\beta(t) =\frac{C_1^{\beta}(t)M_{22}(t)-C^\beta_{2}(t)M_{12}(t)}{\mbox{det}[\hat{M}(t)]},
\ee
\be\label{correct2}
{\Delta\calb^2}^\beta (t)=\frac{C_2^{\beta}(t)M_{11}(t)-C^\beta_{1}(t)M_{21}(t)}{\mbox{det}[\hat{M}(t)]},
\ee
and therefore we can see, from the expressions of the coefficients in Eqs. (\ref{m}) and (\ref{c}), that the dynamic of the system within the adiabatic regime is fully determined by the {\it{frozen equilibrium}} solutions of Eq. (\ref{setfrozen}), since the linear response corrections $\Delta\lambda^{V,\omega}(t)$ and ${\Delta\calb^2}^{V,\omega}(t)$ are evaluated only with the frozen $\lambda^f(t)$ and ${\calb^f}^2(t)$.

\section{Nanomotors}\label{nanomotors}
In this section we apply the formalism developed in Sec. \ref{formalism} to study the performance of {\it{ac}}-driven quantum dots as nanomotors that convert electrical energy into work through the exchange of energy between the {\it{dc}}-source (the battery maintaining the bias voltage $V$) and the {\it{ac}}-sources of the time-dependent driving. 

These processes of energy conversion are described by non-equilibrium thermodynamics in terms of generalized driving forces $X_i$, and the fluxes $J_i$ induced by those applied forces. More specifically, it was recently shown in Ref. \cite{ludovico} that in the case of nanomotors, the relevant forces are $X_1=eV/T$ and $X_2=\hbar\omega/T$, and their respective conjugated variables (or fluxes) are the particle current $J_1=I/e$ entering the reservoir at lower chemical potential, and $J_2=P^{ac}/\hbar\omega$, with $P^{ac}$ being the power (work per unit of time) performed by the {\it{ac}}-potentials. We stress that we are focusing on the performance of the system after a complete cycle of the machine, so that all the fluxes are averaged over one period of the {\it{ac}}-driving.
%(and therefore they are time-independent).  
%MC a little too trivial
\subsection{Linear response and conservation laws}

In the adiabatic regime, and for a small bias voltage, the generalized forces are small and the relationship between fluxes and forces is linear
\be\label{linearresp}
\begin{pmatrix}I/e \\ P^{ac}/\hbar\omega   \end{pmatrix}=\hat{L}\begin{pmatrix}eV/T   \\ \hbar\omega/T  \end{pmatrix},
\ee
where the linear response coefficients collected in the matrix $\hat{L}$ are known as {\it{Onsager}} coefficients, each with a clear physical meaning. The coefficient $L_{11}$ is proportional to the electrical conductance $G$, $L_{12}$ describes the adiabatic quantum pumping of particles due to the {\it{ac}}-driving. On the other hand, the frequency-dependent part of the {\it{ac}}-power is described by $L_{22}$, whereas $L_{21}$ captures its modifications due to the applied bias voltage. The latter, is the coefficient related to the exchange of energy between {\it{dc}} and {\it{ac}} sources, and its sign depends on whether the machine is operating as a motor, or in reverse, as a generator. In this work, we are focusing on the performance as a motor, for which $L_{21}< 0$, meaning that the system is performing work on the {\it{ac}}-sources. Then, we will say that this off-diagonal coefficient is related to the output power $P_{out}$, in the sense that it corresponds to the energy flux that could be later transformed, in a proper setup, into another kind of energy, say mechanical energy. 
%In the case of a generator, $L_{21}>0$, and it will correspond to the portion of the power injected by the {\it{ac}}-sources on the system that is used to pump particles against the voltage drop.
More precisely, the relations between the Onsager elements and the transport coefficients read
\ba\label{transport}
G &=& e^2\frac{L_{11}}{T}, \,\,\,\,\,Q_p =he \frac{L_{12}}{T},\nonumber\\
P_{out} &=& -\frac{L_{21}}{T}\frac{heV}{\tau},\,\,\,\,P^{ac}_{diss}=\frac{L_{22}}{T}\left(\frac{h}{\tau}\right)^2,
\ea 
where $Q_p$ is the pumped charge per driving period $\tau=2\pi/\omega$, and $P^{ac}_{diss}$, as it will be shown later, is the portion of the power developed by the {\it{ac}}-sources that is related to dissipation due to heating of the system. 

Within linear response, the rate of entropy production averaged over one period of the {\it{ac}}-drivings is $\dot{S}=\vec{J}\cdot\vec{X}$, which for a nanomotor reads
\be\label{entropy2}
T\dot{S}=I\cdot V+P^{ac},
\ee
where we can identify $P^{dc}=I\cdot V$ as the power developed by the battery maintaining the bias voltage (i.e the power of the {\it{dc}}-source). From Eqs. (\ref{linearresp}) and (\ref{transport}), it can be seen that in terms of the transport coefficients the power developed by the driving sources are
\be\label{powerdc}
P^{dc}=GV^2+Q_pV/\tau,
\ee
and
\be\label{powerac}
P^{ac}=-P_{out}+P^{ac}_{diss}.
\ee
On the other hand, it was also shown in Ref. \cite{ludovico} that systems with time-reversal symmetry, thus without an applied magnetic field, satisfy the following Onsager reciprocal relation for the crossed coefficients:
\be\label{onsager}
L_{12}=-L_{21}.
\ee
Thus, we can notice from Eq. (\ref{transport}) that the above reciprocal relation leads to Eq. (\ref{power}), $P_{out}=Q_pV/\tau$. The latter is a positive quantity for a motor, and captures the physics of its operation: the conversion between electrical energy into work performed on the {\it{ac}}-sources.   

In irreversible processes, the entropy production is associated with the power dissipated within a cycle, $P_{diss}$, through $T\dot{S}=P_{diss}$, which corresponds to dissipation as heat inside the reservoirs \cite{lowfreq1}. Then, from Eqs. (\ref{entropy2}), (\ref{powerdc}) and (\ref{powerac})
\be\label{diss}
P_{diss}=P^{dc}+P^{ac}=P^{dc}_{diss}+P^{ac}_{diss},
\ee
where we have defined $P^{dc}_{diss}=GV^2$ as the power dissipated by the voltage source.  
The second law of thermodynamics, meaning that $\dot{S}\geq 0$, imposes some conditions on the linear response transport coefficients. In this case, since the conductance is positive defined, the positivity of the entropy production leads to $P^{ac}_{diss}\geq 0$. 

\begin{figure}[h]
   \includegraphics[width=0.4\textwidth]{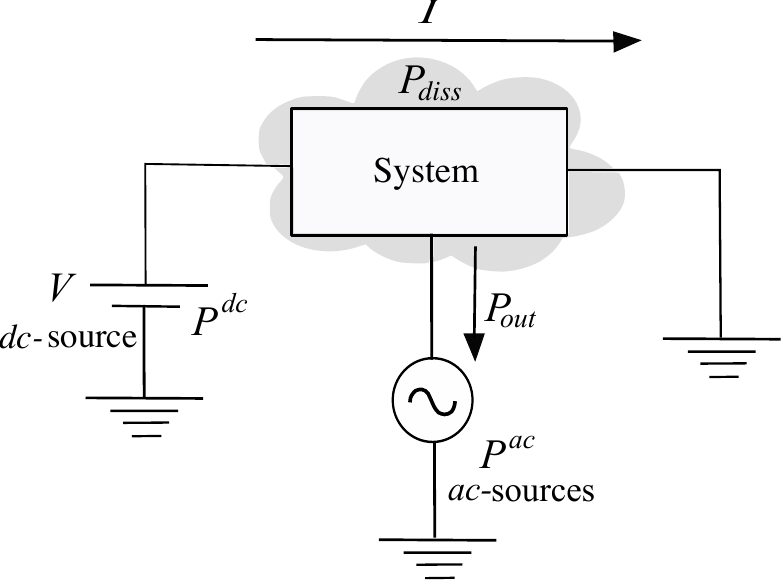}
   \caption{Schematic circuit of the system and the driving sources. The box represents the quantum dot together with the two reservoirs. $P^{dc}$ is the power developed by the battery maintaining the bias voltage $V$ between the reservoirs, while $P^{ac}$ corresponds to the power developed by the {\it{ac}}-sources. As a result of the driving a current $I$ flows through the system. The cloud represents the power that is dissipated as heat inside the reservoirs, $P_{diss}=P^{ac}+P^{dc}$. On the other hand, an energy flux $P_{out}$ is transfered from the {\it{dc}}-source to the {\it{ac}}-sources. }\label{energysketch}
 \end{figure}

The above equation together with Eq. (\ref{power}) constitute the laws for the conservation of the energy in the system. Eq. (\ref{diss}) tells that the total power developed by the driving sources, both {\it{dc}} and {\it{ac}}, must be equal to the net dissipation in the system. Moreover, Eq. (\ref{power}), establishes that part of the power developed by the {\it{dc}} source is transfered to the {\it{ac}} sources through the adiabatic quantum pumping of charge. In each cycle, $Q_p$ changes are pumped into the reservoir at lower chemical potential, and the gain in electrical energy ($Q_pV$) is used to perform work on the {\it{ac}}-sources ($\tau P_{out}$). 
This is illustrated in Fig. {\ref{energysketch}}, as an electronic circuit. The system, which is composed by the quantum dot and the two reservoirs, is represented by a box. In the motor mode, both {\it{dc}} and {\it{ac}} sources drive an electric current $I$ flowing from left to right, in the same direction as the voltage drop. And, as a result of this charge flow, part of the energy injected by the sources is dissipated as heat (schematized as a cloud), while an amount $P_{out}$ is delivered by the system to the {\it{ac}}-sources. 

%%%%%%%%%%
\vspace{1em}
 %%%%%%%%%%
\subsection{Efficiency at maximum power}\label{secefi}
The efficiency of a motor is defined as the ratio between the power performed by the system on the {\it{ac}}-sources, $P^{s\rightarrow ac}=-P^{ac}$, and the power injected by the voltage source $P^{dc}$,
\be\label{eff}
\eta=\frac{P^{s\rightarrow ac}}{P^{dc}}.
\ee
We are interested in analyzing the efficiency of the system at maximum power $P^{s\rightarrow ac}_{max}$. For that, we write the {\it{ac}} power in terms of the fluxes and forces 
\be
P^{s\rightarrow ac}=-TX_2J_2=-T(L_{21}X_1X_2+L_{22}X_2^2).
\ee
Then we can see that, as a function of $X_2$, it is maximimum for
\be\label{xmax}
X_2=-\frac{L_{21}}{2L_{22}}X_1,
\ee
taking the value
\be
P^{s\rightarrow ac}_{max}=\frac{L_{21}^2}{4L_{22}}X_1^2T=\frac{P_{out}^2}{4P^{ac}_{diss}}.
\ee
Thus, the maximum power that can be performed on the {\it{ac}}-sources is dictated by the relation between the power $P_{out}$ transfered from the voltage source to the {\it{ac}}-sources, and the power $P_{diss}^{ac}$ that is dissipated by the {\it{ac}}-sources. The efficiency at maximum power can be obtained by evaluating Eq. (\ref{eff}) at the relation (\ref{xmax}).
For systems obeying the Onsager's relation (\ref{onsager}), it reads  
\be\label{efficiency}
\eta(P_{max}^{s\rightarrow ac})=\frac{1}{2}\frac{\xi}{\xi+2},
\ee
with
\be
\xi=\frac{L_{12}^2}{L_{11}L_{22}}=\frac{P_{out}^2}{P^{dc}_{diss}P^{ac}_{diss}},
\ee
being the figure of merit introduced in Ref. \cite{ludovico}. The efficiency in Eq. (\ref{efficiency}) is a monotonically increasing function of $\xi$, with $\eta(P^{s\rightarrow ac}_{max})=0$ when $\xi=0$ and $\eta(P^{s\rightarrow ac}_{max})\rightarrow 1/2$ for $\xi\rightarrow \infty$. In this way, large values of the figure of merit are needed to achieve high efficiencies, which in turn requires a large pumping of charge coefficient $L_{12}$ (high values for $P_{out}=Q_pV/\tau$) along with a small product $L_{11}L_{12}$ (low dissipation of the {\it{dc}}-source and/or the {\it{ac}}-sources).

\subsection{Transport coefficients}

The coefficients of the Onsager matrix $\hat{L}$ can be written, within the non-equilibrium Green's function formalism, in terms of the Green's functions of the central quantum dot \cite{ludovico, dyson} in Eq. (\ref{greensdot}). For that, we have to compute the fluxes $J_1=I/e$ and $J_2=P^{ac}/\hbar\omega$, and collect the terms which are first order in the small forces $X_1=eV/T$ and $X_2=\hbar\omega/T$.

The charge flux $I$ flowing through the system can be computed in terms of the current of particles entering the reservoirs (see Eq. (\ref{particlecurrent})) as $I/e=\overline{\dot{N}_R(t)}=-\overline{\dot{N}_L(t)}$, where the line on top denotes the temporal average over a period $\tau$ of the {\it{ac}}-drivings: $\overline{O}=\int^\tau_0 O(t)/\tau$. We have also used the fact that the time-dependent MF Hamiltonian in Eq. (\ref{hammf}) conserves the charge (and the energy). In this way, the time-averaged particle flux leaving the left reservoir must be equal to the flux entering the right one, since no net charge can be stored in the quantum dot.   

Now, by following the steps detailed in Appendix \ref{appendix1} for slow driving, the resulting flux reads
\ba\label{charge1}
I & = & \frac{2e}{\hbar}\!\int^\tau_0\!\!\frac{dt}{\tau}\!\!\int\frac{d\ve}{\pi} \left[\tilde{\Gamma}_R(t) \mbox{Im}\left\{G^r(t, \ve)f_R(\ve)+\frac{G^<(t,\ve)}{2}\right\}\right.\nonumber\\
&&\left. +\frac{\hbar}{2}\dot{\tilde{\Gamma}}_R(t) \mbox{Re}\left\{\partial_\ve G^r(t,\ve)\right\}f_R(\ve)\right].
\ea

On the other hand, the power developed by the {\it{ac}}-sources is defined as 
$P^{ac}=\overline{\langle\partial_t H^{MF}_{SB}\rangle }$. After using equations (\ref{num}) and (\ref{realpart}), it is easy to show that the derivative in time of the MF Hamiltonian in (\ref{hammf}) reads
\ba
\langle\partial_t H^{MF}_{SB}\rangle & = & \dot{{\ve}}_{d}(t) N_d(t)+\!\!\!\sum_{\alpha,k_\alpha ,\sigma}\!\!2\mbox{Re}\left\{\dot{{w}}_\alpha(t)\calb(t)\langle c^\dagger _{k_\alpha ,\sigma} f_\sigma\rangle\right\},\nonumber\\
\ea
from which we can see that, as expected, the work %MC I used work here instead of power
is performed on the system due to the change in time of the applied {\it{ac}}-potentials, $\dot{\ve_d}(t)$ and $\dot{w}_\alpha(t)$, while the variation in time of the additional {\it{ac}}-parameters accounting for the interactions $\dot{\lambda}(t)$ and $\dot{\calb}(t)$ indirectly contributes through the evolution of the expectation values $\dot{N}_d(t)$ and $\langle c^\dagger_{k_\alpha,\sigma}f_\sigma \rangle(t,t)$. Similarly as for the charge current $I$ (see Appendix \ref{appendix1} for details), we find that the power can be written in terms of the Green's functions of the dot as 
\ba\label{poweracgreen}
P^{ac} & = &\sum_\alpha\int^\tau_0\!\!\frac{dt}{\tau}\!\!\int\frac{d\ve}{\pi} \left[\dot{{\Gamma}}_\alpha(t)\calb^2(t) \mbox{Re}\left\{G^r(t, \ve)\right\}f_\alpha(\ve)\right.\nonumber\\
&&\left. -\frac{\hbar}{2}\frac{\dot{\Gamma}_\alpha(t)\dot{\tilde{\Gamma}}_\alpha(t)}{\Gamma_\alpha(t)}\mbox{Im}\left\{\partial_\ve G^r(t,\ve)\right\}f_\alpha(\ve)\right.\\
&&+\left.\mbox{Im}\left\{\left(\dot{\ve}_d(t)+i\frac{\dot{\Gamma}_\alpha(t)\calb^2(t)}{2}\right)G^<(t,\ve)\right\}\right].\nonumber
\ea
In order to find the linear-response Onsager coefficients $L_{11}$ and $L_{12}$, we need to collect the terms in the current $I$ which are first order in $eV$ and $\hbar\omega$. Nevertheless, in order to express
$J_2=P^{ac}/\hbar\omega$ within linear response and compute $L_{21}$ and $L_{22}$, it is necessary to keep the second order terms in $P^{ac}$ that are proportional to $(eV\hbar\omega)$ and $(\hbar\omega)^2$. We note that $P^{ac}$ has no linear or quadratic terms in $eV$ since no {\it{ac}}-power can be generated in the static situation ($\omega=0$). Moreover, its linear term in $\hbar\omega$ is zero because it is related to the change of the entropy in reversible processes, which vanishes under a cycle that begins and ends at the same equilibrium state \cite{lowfreq1}. 

Similarly as in Section \ref{formalism}, we expand the retarded and lesser Green's functions in Eqs. (\ref{charge1}) and (\ref{poweracgreen}) up to first order in the change of the {\it{ac}}-parameters, $\dot{\tilde{\ve}}_d(t)$ and $\dot{\tilde{\Gamma}}_\alpha(t)$ (see Appendix \ref{lowfreq}). Besides, we combine this treatment with an expansion of the Fermi functions of the reservoirs $f_\alpha(\ve)$ in powers of $eV$. In this way, we find that
\be\label{l11}
L_{11}=-\frac{T}{\hbar\pi}\!\!\int^{\tau}_0\!\!\frac{dt}{\tau}\!\!\int d\ve\, \partial_\ve f\frac{\tilde{\Gamma}_L^f(t)\tilde{\Gamma}_R^f(t)}{\tilde{\Gamma}^f(t)}\rho^f(t,\ve),
\ee
and
% \be\label{l12}
% L_{12}=-\frac{T}{h\pi}\int^{\tau}_{0}dt\int {d\ve}\frac{df}{d\ve}
% \frac{\rho^f(t,\ve)}{\tilde{\Gamma}^f(t)} \left[\dot{\tilde{\ve}}^f_{d}(t){\tilde{\Gamma}_R^f(t)}+ \dot{\tilde{\Gamma}}_R^f(t)(\ve-\tilde{\ve_{d}}^f(t))\right].
% \ee
\be\label{l12}
L_{12}=\frac{T}{h\pi}\!\!\int^{\tau}_{0}\!\!dt\!\!\int\!\!{d\ve}\,\partial_\ve f\rho^f(t,\ve)
\frac{{{\tilde{\Gamma}^{f^2}_R}}(t)}{\tilde{\Gamma}^f(t)}\partial_t\left(\frac{(\ve-\tilde{\ve}_{d}^f(t))}{\tilde{\Gamma}_R^f}\right).
\ee

In the case of $P^{ac}$, since it is a second-order quantity, it has a contribution not only of the {\it{frozen}} solutions but also of the corrections $\Delta\lambda^{V,\omega}(t)$ and ${\Delta\calb^2}^{V,\omega}(t)$. Explicit expressions of the Onsager coefficients $L_{21}$ and $L_{22}$ in terms of the frozen solutions and the corrections can be found in Appendix \ref{appendix3}, where we also show the validity of the Onsager reciprocal relation (\ref{onsager}), and that the dissipative coefficient reads  
\ba\label{l22}
L_{22}&=&-\frac{T}{h2\pi\omega}\!\!\int^\tau_0\!\!dt\!\!\int\!\!d\ve\,\partial_\ve f\!\!\left\{\!\!\left[\!{\tilde{\Gamma}}^f(t)\rho^f(t,\ve)\partial_t\!\!\left(\frac{\ve-\tilde{\ve}_{d}^f(t)}{\tilde{\Gamma}^f(t)}\right)\!\!\right]^2\right.\nonumber\\ 
&&\left.+\rho^f(t,\ve)\left(\sum_\alpha \frac{(\dot{\tilde{\Gamma}}^f_\alpha(t))^2}{2\tilde{\Gamma}^f_\alpha(t)}-\frac{(\dot{\tilde{\Gamma}}^f(t))^2}{2\tilde{\Gamma}^f(t)}\right)\right\}.
\ea

Therefore, by using the explicit expressions for the corrections in (\ref{correct1}) and (\ref{correct2}), we demonstrate that all the linear response coefficients are evaluated only at the {\it{frozen}} equilibrium solutions of Eq. (\ref{setfrozen}).  
Furthermore, we find that Eqs. (\ref{l11}), (\ref{l12}) and (\ref{l22}) are the same as those obtained for the non-interacting quantum dot (by evaluating the Hamiltonian (\ref{hamdot}) at $U=0$) with the substitutions $\Gamma_\alpha(t)\rightarrow \tilde{\Gamma}_\alpha^f(t)$ and $\ve_d(t)\rightarrow \tilde{\ve}_d^f(t)$. 

\section{Results}\label{Results}
As an illustrative example, we consider the case $\ve_d(t)=0$, with hybridizations $\Gamma_L(t)=\Gamma^{dc}+\Gamma^{ac}\cos(\omega t)$ and $\Gamma_R(t)=\Gamma^{dc}-\Gamma^{ac}\sin(\omega t)$. The hybridization with the reservoirs must oscillate with a relative phase lag, since at least two different time-dependent parameters are needed in order to have quantum pumping within the adiabatic regime \cite{brouwen}. We also focus on the situation in which the reservoirs are at the same temperature $T=0$.

\begin{figure}[h]
  \includegraphics[width=0.47\textwidth]{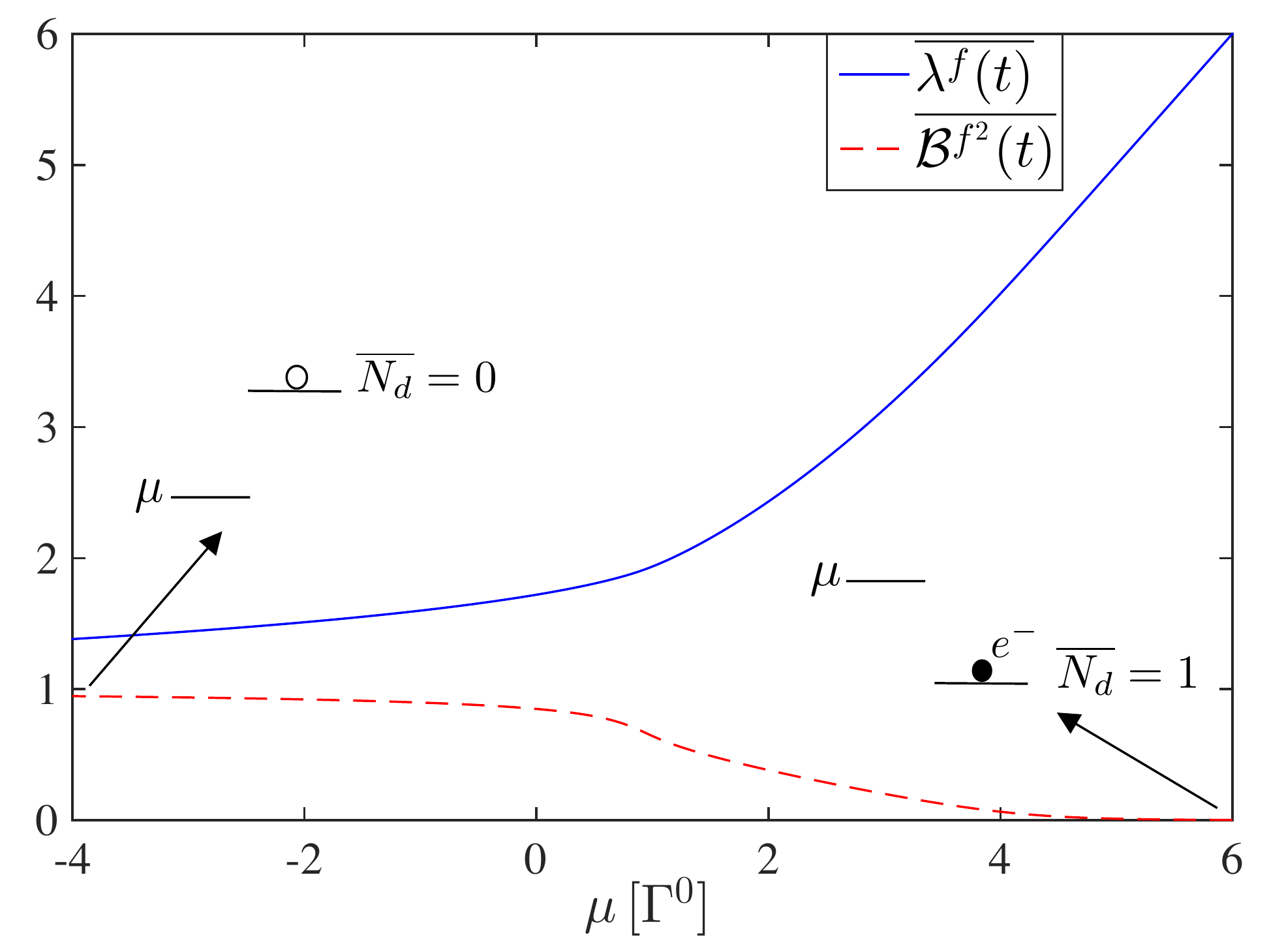}
  \caption{Time average over one driving period of the frozen Lagrange multiplier $\overline{\lambda^f}(t)$ and density of holes $\overline{\calb^{f^2}}(t)$, as a function of the chemical potential of the reservoirs $\mu$. The {\it{ac}}-parameters $\lambda^f(t)$ and $\calb^{f^2}(t)$ are computed by solving the equilibrium-like non-linear system of equations in Eq. (\ref{setfrozen}) at every frozen time $t$, considering a zero bias voltage $eV=0$ applied between the reservoirs. Since in this example $\ve_d=0$, then positive values of the chemical potential $\mu$ correspond to the energy level of the dot being below the Fermi energy, while the opposite situation occurs when $\mu<0$. The parameters of the driving are, $\Gamma^{dc}=\Gamma^{ac}=0.5$, and $\hbar\omega=0.001$. The temperature of the reservoirs is $T=0$. All the energies are in units of the {\it{dc}} component of the total hybridization, $\Gamma^0\equiv\overline{\Gamma(t)}=2\Gamma^{dc}$.}\label{fig2}
\end{figure}

As was shown in the previous section, the Onsager coefficients describing the linear-response charge and energy transport through the driven interacting quantum dot (see Eqs. (\ref{l11}), (\ref{l12}) and (\ref{l22})), are evaluated only at the {\it{frozen}} parameters $\lambda^f(t)$ and $\calb^{f^2}(t)$. The latter are found by solving the non-linear system of equations in (\ref{setfrozen}) at every {\it{frozen}} time $t$. We emphasize again that in this {\it{frozen picture}} the system is considered to be at equilibrium at every time $t$ as in a sequence of snapshots, so that the variable $t$ is treated as a parameter. 
In Fig. \ref{fig2} we show the average value over one driving period of the Lagrange multiplier $\overline{\lambda^f(t)}$, together with the bosonic field $\overline{\calb^{f^2}(t)}$ as a function of the chemical potential $\mu$ of both of the reservoirs (since in Eq. (\ref{setfrozen}) the bias is $eV=0$). We can see that when the level of the dot is deep below the chemical potential ($\mu\gg0$), the average density of holes vanishes $\overline{\calb^{f^2}(t)}\rightarrow 0$, which means that the level is occupied with only one electron. Moreover, in the same limit, the effective energy of the level is in resonance with the chemical potential $\overline{\tilde{\ve}_d^f(t)}=\overline{\lambda^f(t)}\sim \mu$, as expected in the Kondo regime. In the opposite situation when the dot has an energy far above the Fermi level $\mu\ll 0$, the field $\overline{\calb^{f^2}(t)}\sim 1$ so that the quantum dot is empty.  

%%%%%%%%%%
\begin{figure}[h!]
  \includegraphics[width=0.48
  \textwidth]{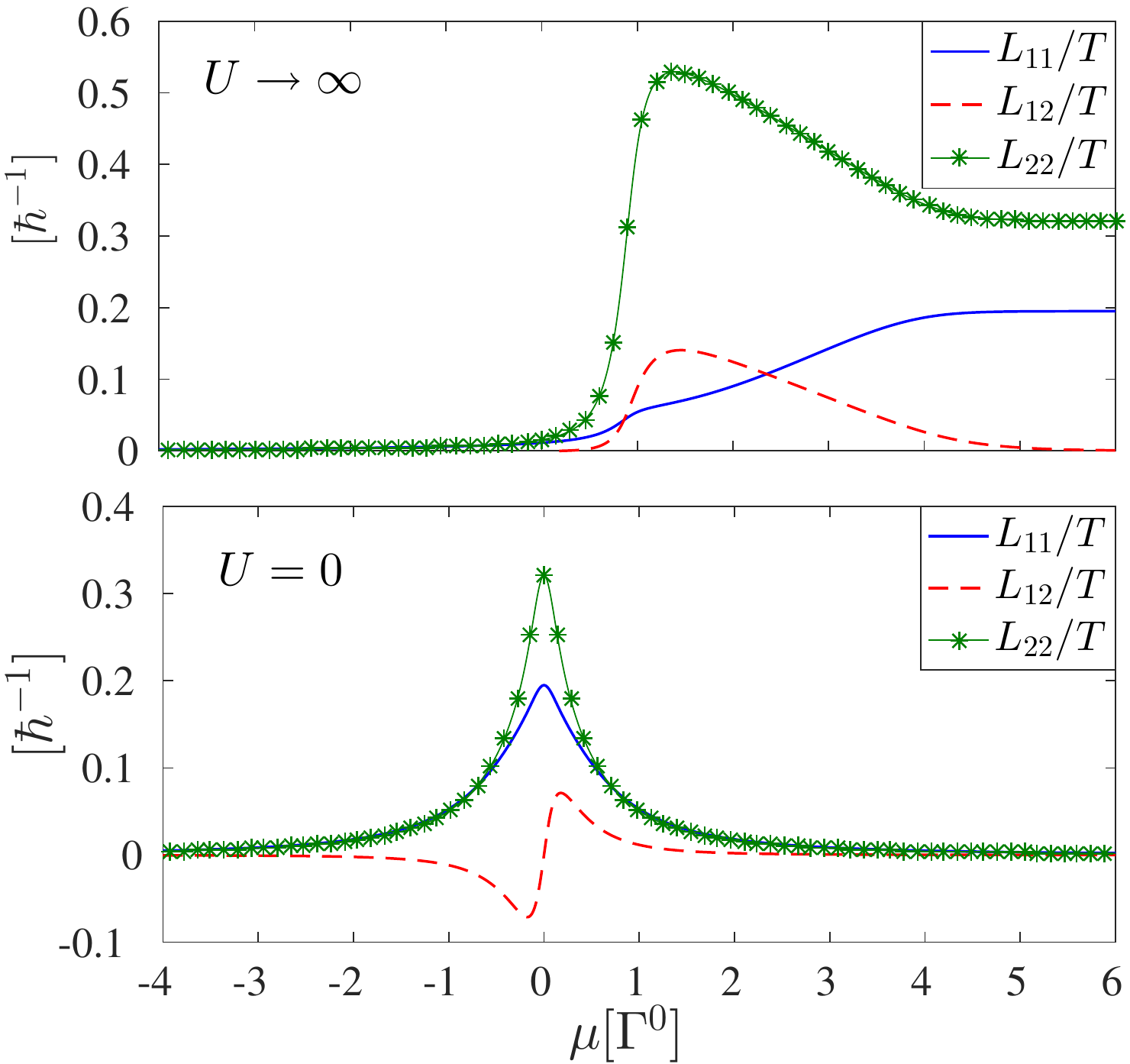}
  \caption{Onsager coefficients as a function of the chemical potential $\mu$ of the reservoir on the left. Top panel corresponds to the strongly interacting case $U\rightarrow \infty$, while the bottom panel shows the results for non-interacting electrons $U=0$. All the parameters are the same as in Fig. \ref{fig2}. }\label{fig3}
\end{figure}

%%%%%%%%%%
Now, we turn to  the behavior of the linear response transport coefficients in Eqs. (\ref{l11}), (\ref{l12}) and (\ref{l22}) as functions of the chemical potential. The results can be found in Fiq. \ref{fig3}, where for comparison we also slow (bottom panel) the same coefficients for non-interacting electrons. In the $U=0$ limit, the conductance $G=e^2L_{11}/T$ reaches its maximum value when the chemical potential is in resonance with the energy level of the dot (i.e. $\mu=0$). 
Driving the system with two barriers oscillating with a phase lag $\delta=\pi/2$, decreases the conductance and favors pumping by dynamically putting the system off resonance. Therefore, the pumping coefficient $L_{12}$ vanishes for $\mu=0$ and attains its peaks when the chemical potentials is apart from the resonance \cite{ludovico, Splett}. 

\begin{figure}[h]
  \includegraphics[width=0.48\textwidth]{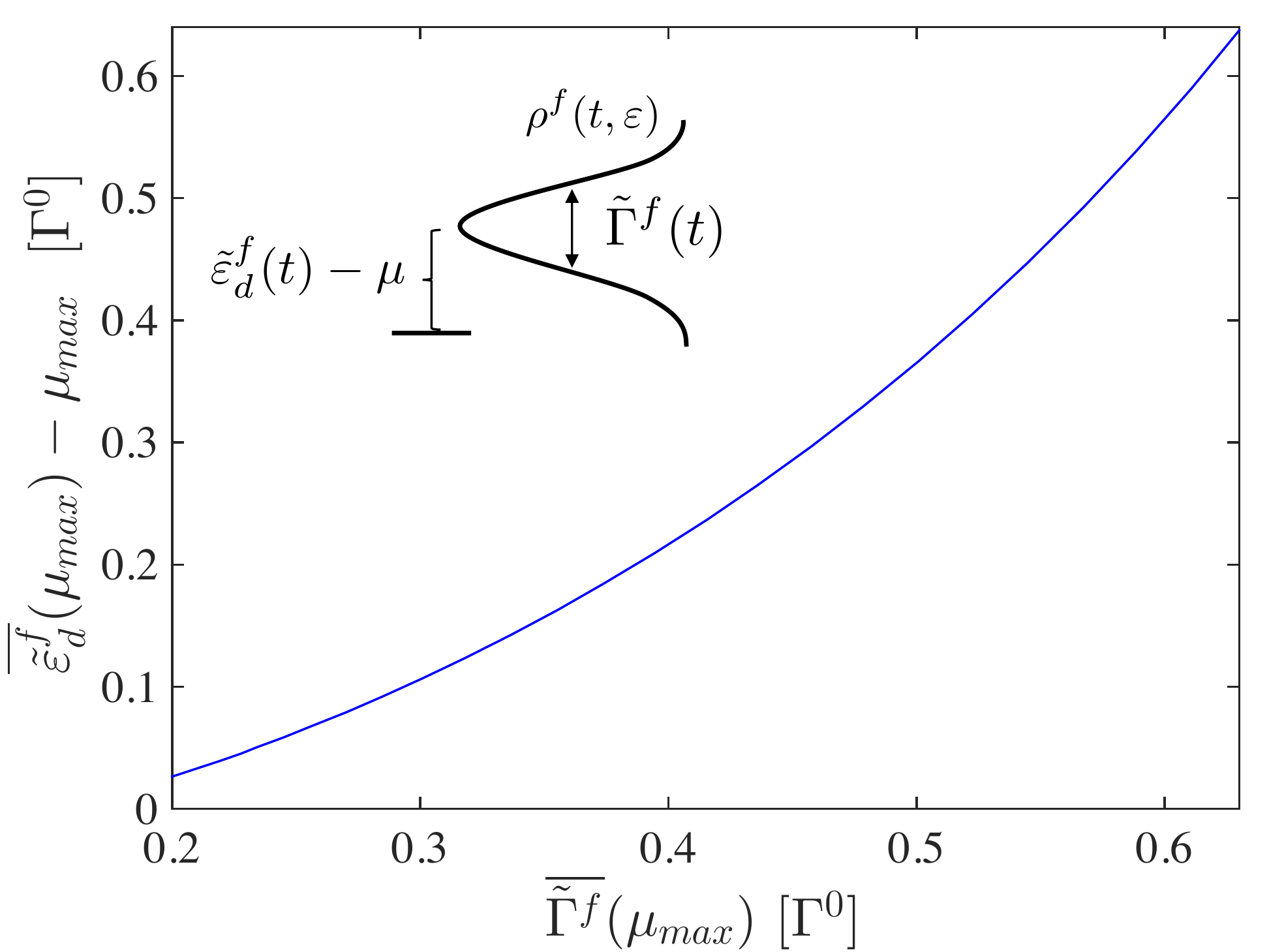}
  \caption{Time average of the energy difference $\tilde{\ve}^f_d(t)-\mu$ vs the averaged effective total hybridization $\overline{\tilde{\Gamma}^f}$, evaluated at the chemical potential $\mu_{max}$ for which the pumping coefficient $L_{12}$ achieves its maximum value. The {\it{dc}} component is $\Gamma^{dc}=0.5\Gamma^0$, while the parameter $\Gamma^{ac}$ was varied from $10^{-2}\Gamma^0$ to $\Gamma^{dc}$.}\label{fig4}
\end{figure}

In the strongly interacting limit $U\rightarrow\infty$, the extra {\it{ac}}-parameters, $\lambda^f(t)$ and $\calb^{f^2}(t)$, are introduced through a renormalized hybridization with the reservoirs, $\Gamma_\alpha(t)\calb^{f^2}(t)$, and an effective energy level $\tilde{\ve}^f_d(t)=\lambda^f(t)$. As can be seen from Fig. \ref{fig2} and the top panel in the top panel of Fig. \ref{fig3}, the Lagrange multiplier $\lambda^{f}(t)$ has the effect of moving the resonance from its non-interacting value $\mu=0$, to energies far above the dot level $\mu\gg 0$ (Kondo peak of the conductance \cite{dong}). However the extra {\it{ac}}-parameters introduced by the interactions between the electrons in the dot do not only shift the resonance (as in {\it{dc}}-transport), but also contribute to the pumping of particles. 
Remarkably we find that for some values of $\mu$ the pumping coefficient $L_{12}$ is enhanced with respect to the non-interacting problem, and it is also bigger than the conductance $L_{12}>L_{11}$. This different behavior arises from the new time-dependent quantities introduced by interactions. As in the non-interacting problem, the pumping coefficient vanishes at the resonance, which in this case occurs when $\mu\gg 0$. This can be easily proved by taking at $T=0$ the limit $\lambda^f\rightarrow \mu$ and $\calb^{f^2}\rightarrow 0$ in Eq. (\ref{l12}). On the other hand, $L_{12}$ decays as $\vert\tilde{\ve}_d^f-\mu\vert^{-1}$ when $\mu\ll 0$.
To analyze in more depth the behavior of the pumping coefficient, we focus on the energy $\mu=\mu_{max}$ at which $L_{12}$ reaches its maximum value. Then, we study the relation between the energy distance $(\tilde{\ve}_d^f(t)-\mu_{max})$ and the width of the frozen density of states, which is proportional to the total hybridization $\tilde{\Gamma}^f(t)$. The results are shown in Fig. \ref{fig4}. We find that the maximum of the time-averaged pumping coefficient occurs when the two quantities become comparable, i.e. $(\overline{\tilde{\ve}_d^f}(\mu_{max})-\mu_{max})\sim \overline{\tilde{\Gamma}^f}(\mu_{max})$, so that the effect of the modulation of the hybridizations can be felt. A similar behavior of the pumping coefficient was reported in Refs. \cite{citro, Splett}. 

%%%%%%%%%%
%%%%%%%%%%
Note that in the case with $U=0$, the pumping coefficient changes sign as $\mu$ passes the resonance ($\mu=0$). This means that the system switches from the motor mode with $L_{12}>0$ (when $\mu>0$) to the the generator mode with $L_{12}<0$ (for $\mu<0$). However, in the limit $U\rightarrow\infty$, the pumping of particles vanishes for $\mu<0$, and therefore the system operates only as a motor.

Finally, we turn to the dissipative coefficient $L_{22}$. We can see from Eq. (\ref{l22}) that it has two contributions. The first one, as for the pumping coefficient in Eq. (\ref{l12}), results in a peak at $\mu=\mu_{max}$ and vanishes for $\mu\gg0$, while the second one leads to a finite value at the Kondo peak (see top panel of Fig. \ref{fig3}). We can also observe a significant increase of the maximum value of dissipation (around $L_{22}^{max}\sim 0.55 \hbar^{-1}$) with respect to the non-interacting problem ($L_{22}^{max}\sim 0.32 \hbar^{-1}$). An increment of the dissipation in nanomotors due to electronic interactions was also reported in \cite{bruch}. This effect can be traced back to the fact that there are more {\it{ac}}-parameters in the interacting problem, that contribute to the pumping of charge as well as to the dissipation.

%%%%%%%%%%
\begin{figure}[h]
  \includegraphics[width=0.46\textwidth]{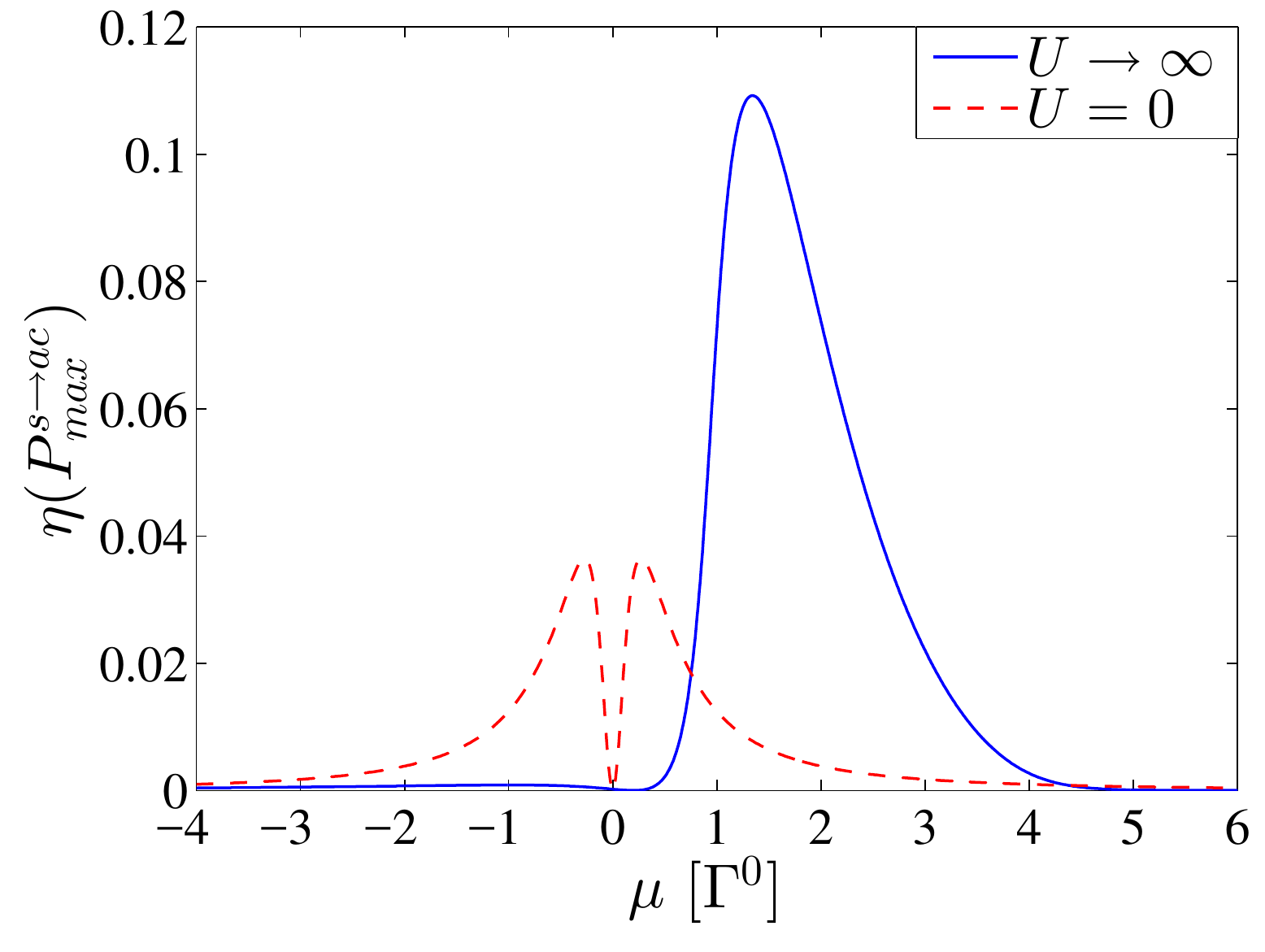}
  \caption{Efficiency at maximum power $\eta(P^{s\rightarrow ac}_{max})$ as a function of the chemical potential. All the parameters are the same as in Fig. \ref{fig2}.}\label{fig5}
\end{figure}

\begin{figure}[h]
  \includegraphics[width=0.49\textwidth]{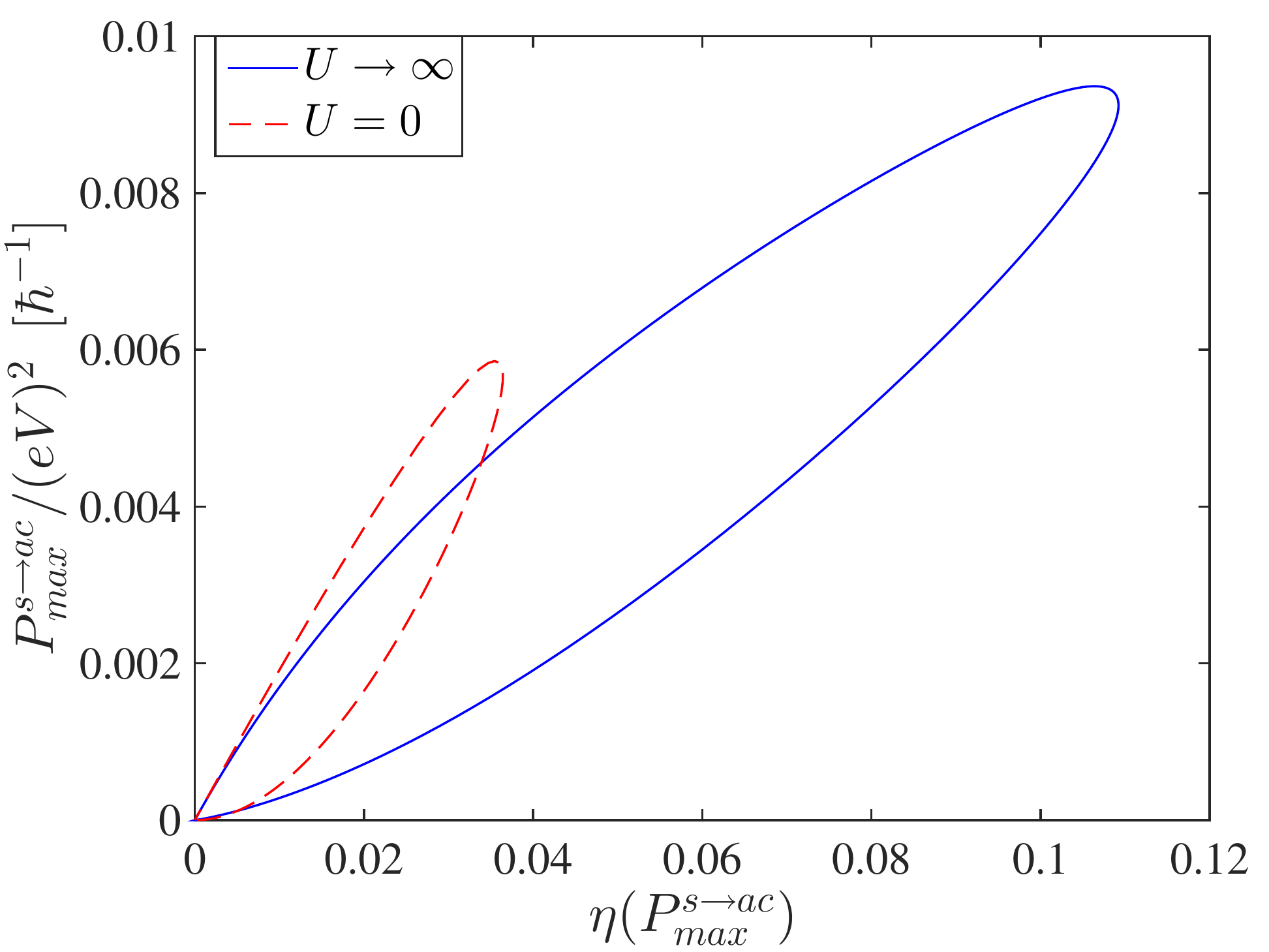}
  \caption{Maximum power performed by the electronic system on the {\it{ac}}-sources $P^{s\rightarrow ac}_{max}/(eV)^2$, as a function of the efficiency $\eta(P^{s\rightarrow ac}_{max})$. All the parameters are the same as in Fig. \ref{fig2}. }\label{fig6}
\end{figure}

%%%%%%%%%%
As discussed in Sec. \ref{secefi}, the enhancement of the pumping effect along with the reduction of the product $L_{11}L_{22}$ favors the improvement of the efficiency in Eq. (\ref{efficiency}). This is why, as we can see in Fig. \ref{fig5}, higher efficiencies can be attained for $U\rightarrow\infty$ in comparison with the non-interacting problem. The improvement of the performance occurs for energies around $\mu_{max}$, for which the pumping coefficient is maximized, and its maximum value is around three times larger than the one obtained for $U=0$ ($\eta^{U\rightarrow\infty}_{max}\sim 3\eta^{U=0}_{max}$). Moreover, not only an enhance of the efficiency can be obtained due to Coulomb interactions, but also the maximum power done by the system on the {\it{ac}}-sources is larger. This is illustrated in Fig. \ref{fig6}, where we show the maximum power divided by the square of the affinity $eV$ (or force) driving the motor, $P^{s\rightarrow ac}_{max}/(eV)^2$, as a function of the efficiency at which the latter power is delivered by the system $\eta(P^{s\rightarrow ac}_{max})$. It can be noticed that, in the strongly interacting limit, the maximum power that is done by the system is higher, and it is delivered more efficiently.

%%%%%%%%%%
%%%%%%%%%%

%%%%%%%%%%
%%%%%%%%%%

\section{Conclusions}\label{conclusions}

In this work we studied the effects of electron-electron interactions on the performance of a quantum-dot-based nanomotor. To address this problem, we considered the simplest  meaningful setup consisting in a two-terminal device with an interacting quantum dot in the middle, as it is illustrated in Fig. \ref{fig1}. Charge and energy transport through the system is driven by a voltage difference between the reservoirs, along with the application of {\it{ac}}-potentials to control the transparency of the tunneling barriers and vary the energy level of the dot. 

We focused on the strongly-interacting limit, and on the adiabatic response regime for which the {\it{ac}}-driving potentials slowly evolve in time. In this context, we developed a method to describe analytically the interacting quantum dot that combines the time-dependent slave-boson mean-field theory with a linear response treatment in the small rate of change of the {\it{ac}}-parameters of the MF Hamiltonian. The advantage and beauty of the formalism we presented here is that the dynamics of the system turns out to be simply described in terms of {\it{frozen}} equilibrium solutions at every time.  Moreover,  the approach is not restricted to the strongly interacting limit or the linear response regime, since the slave-boson approach can also be implemented in the finite-$U$ case using for example the Kotliar-Ruckenstein formalism, and one can keep higher order contributions in the adiabatic parameter. 

In order to study the performance of our system as a motor, we analyzed the relation between the charge current flowing through the quantum dot and the power developed by the {\it{ac}}-driving sources. We computed the relevant transport coefficients and we showed that they satisfy Onsager reciprocity relations.  We found that, similarly to the stationary case, all the linear response transport coefficients are obtained from the expressions for a non-interacting quantum dot with correlation-induced renormalizations of the dot energy level and hybridizations with the reservoirs.

Finally, as an illustrative example, we considered the system being at zero temperature and with a constant energy level of the dot. We found that the additional {\it{ac}}-parameters introduced by the interactions, due to the temporal dependence of the Lagrange multiplier and the bosonic field, lead to two main effects. One is the shift of the resonance from its non-interacting value to energies deep below the Fermi level (Kondo peak), and the second one is the enhancement of the efficiency with respect to a non-interacting dot. The latter can be understood from the fact that the extra {\it{ac}}-parameters accounting for the interactions increase the pumping of particles while decreasing the electrical conductance.

%\newpage
\section{Acknowledgements} 
We acknowledge support from the H2020 Framework Programme under ERC Advanced Grant No. 692670 “FIRSTORM’, The Ministero dell’Istruzione Universit`'a e Ricerca through PRIN 2015 (Prot. 2015C5SEJJ001) and SISSA/CNR project ”Superconductivity, Ferroelectricity and Magnetism in bad metals” (Prot. 232/2015).

\appendix
\section{Non-equilibrium Green's function formalism and linear response approximation}\label{appendix1}

The expectation value $\langle c^\dagger_{k_\alpha\sigma}f_\sigma\rangle(t,t)$ is a Green's function that involves operators of the reservoirs as well as from the dot. By solving the Dyson equation and using Langreth rules \cite{langreth}, the above function can be expressed as follows 
\ba\label{ap11}
\sum_{k_\alpha\in\alpha}\langle c^\dagger_{k_\alpha\sigma}f_\sigma\rangle(t,t)\!&=&\!-i \!\!\int\!\!dt'\tilde{w}^*_{\alpha}(t')\!\left(\!G^r_\sigma(t,t')g_{\alpha,\sigma}^<(t'\!-t)\right.\nonumber\\
&&+G^<_\sigma(t,t')g_{\alpha,\sigma}^a(t'-t)\Big),
\ea
with $\tilde{w}^*_{\alpha}(t')=w^*_{\alpha}(t')\calb(t')$, and $G^r_\sigma(t,t')=-i\theta(t-t')\langle\{f_\sigma(t)f^\dagger_\sigma (t')\}\rangle$ and $G^<_\sigma(t,t')=-i\langle f_\sigma^\dagger(t')f_\sigma (t)\rangle$ being respectively the retarded and the lesser Green's functions of the quantum dot connected to the leads. On the other hand, within the wide band limit, the corresponding Green's functions of the uncoupled reservoirs are
\ba\label{greenres}
g^<_{\alpha,\sigma}(t-t')& = &i\,\rho_{\alpha}\int\frac{d\ve}{2\pi}f_\alpha(\ve)e^{-i\frac{\ve}{\hbar}(t-t')}\\
g^{a}_{\alpha,\sigma}(t-t')& = &i\delta(t-t')\frac{\rho_\alpha}{2},
\ea
where $\rho_{\alpha}$ is the constant density of states of the reservoir $\alpha$ and $f_\alpha(\ve)$ corresponds to its the Fermi-Dirac distribution.

Now, for slow {\it{ac}}-driving, we perform the following linear approximation in the temporal variation of the tunneling elements
\be
\tilde{w}^*_{\alpha}(t')\sim \tilde{w}_{\alpha}^*(t)+\frac{d\tilde{w}^*_{\alpha}(t)}{dt}(t'\!-t).
\ee 
In this way, Eq. (\ref{ap11}) reads
\ba
\sum_{k_\alpha\in\alpha}\!\!\langle c^\dagger_{k_\alpha\sigma}f_\sigma\rangle(t,t) & = & \tilde{w}_\alpha^*(t)\rho_\alpha\!\!\int\!\!\frac{d\ve}{2\pi} \!\left(\!\!G^r_\sigma(t, \ve)f_\alpha(\ve)\!+\!\!\frac{G^<_\sigma(t,\ve)}{2}\!\!\right)\nonumber\\
&&+i{\hbar}\rho_\alpha\frac{d\tilde{w}^*_{\alpha}(t)}{dt}\!\!\int\!\!\frac{d\ve}{2\pi}\partial_\ve G^r_\sigma(t,\ve)f_\alpha(\ve),
\ea
with $G^r_\sigma(t,\ve)$ and $G^<_\sigma(t,\ve)$ being the Fourier transforms of the Green's functions of the dot, which are defined in Eq. (\ref{greensdot}).
\section{Low frequency expansion}\label{lowfreq}
In the slow driving regime, for which the typical driving frequency of the {\it{ac}}-fields is small ($\omega\rightarrow 0$), an exact analysis up to linear order in $\omega$ can be done by expanding the Green functions $G^{r}(t,\ve)$ and $G^{<}(t,\ve)$ up to first order in the temporal variation of {\it{ac}}-parameters of the Hamiltonian of the full system \cite{lowfreq1,lowfreq2,ludoarr}.
In the case of the MF Hamiltonian in Eq. (\ref{hammf}), the Green's functions of the dot read
\be
G^{r}(t,\ve)=G_0(t,\ve)+i\hbar \partial_\ve G_0(t,\ve)\!\!\left[\dot{\tilde{\ve}}_d-i\frac{\dot{\tilde{\Gamma}}}{2}\right]\!\! G_0(t,\ve),
\ee
and
\ba
G^<(t,\ve) &=&G^{<}_0(t,\ve)\\
&&+i\hbar\,\dot{\tilde{\ve}}_d\left[\partial_\ve G_0(t,\ve)G^{<}_0(t,\ve)\!+\!\partial_\ve G^{<}_0(t,\ve) G^{\dagger}_0(t,\ve)\right]\nonumber\\
&& +\hbar\,\frac{\dot{\tilde{\Gamma}}}{2}\left[\partial_\ve G_0(t,\ve) G^{<}_0(t,\ve)\!-\!\partial_\ve G^{<}_0(t,\ve) G^{\dagger}_0(t,\ve)\right]\nonumber\\
&& +i{\hbar}\vert G_0(t,\ve)\vert^2 \frac{\partial_\ve\partial_t\tilde{\Sigma}^<(t,\ve)}{2}\nonumber\\
&&+i\hbar \partial_\ve G_0(t,\ve)G_0^{\dagger}(t,\ve)\partial_t\tilde{\Sigma}^<(t,\ve),\nonumber
\ea
with
\be\label{retfrozen2}
G_0(t,\ve)=\left[\ve-\tilde{\ve}_d(t)+i\frac{\tilde{\Gamma}(t)}{2}\right]^{-1},
\ee
being the retarded Green function describing the regime in which the electrons instantaneously adjust its potential to the {\it{ac}}-fields. The lesser Green function is defined as $G^{<}_0(t,\ve)=\vert G_0(t,\ve)\vert\tilde{\Sigma}^<(t,\ve)$, with $\tilde{\Sigma}^<(t,\ve)=i\sum_{\alpha=L,R}f_\alpha(\ve)\tilde{\Gamma}_\alpha(t)$.

\section{Linear response coefficients of the flux $J_2$ and validity of the Onsager's relation}\label{appendix3}
By keeping the terms in Eq. (\ref{poweracgreen}) which are proportional to $eV\hbar\omega$, we find that
\ba\label{l21}
L_{21}&=&\frac{T}{h}\!\!\int^{\tau}_{0}\!\!\!dt\!\left\{\!\int\!\!\frac{d\ve}{\pi}\partial_\ve f\frac{\rho^f}{{\Gamma}}\!\!\left[\dot{\ve}_d{{\Gamma}_R}+\dot{{\Gamma}}_R(\ve-\tilde{\ve}_{d}^f)\right]\right.\nonumber\\
&&+\frac{\dot{\ve}_d}{eV}\left[C_2^V-\Delta{\calb^2}^{V}\right]\\
&&\left.+\frac{\dot{\Gamma}\calb^{f^2}}{eV\Gamma}\!\left[C_1^V-\Delta\lambda^{V}-\lambda^f\frac{\Delta{\calb^2}^{V}}{\calb^{f^2}}\right]\right\}.\nonumber
\ea
On the other hand, from the term $\propto (\hbar\omega)^2$ we get
\ba\label{l22sup}
L_{22}&=&-\frac{T}{h\omega}\!\!\int^\tau_0\!\!dt\!\int\!\!\frac{d\ve}{2\pi}\partial_\ve f\rho^f\Bigg\{ \rho^f\dot{\ve}_d\,\dot{\tilde{\ve}}_{d}^f\nonumber\\
&&\left.+\frac{\rho^f}{\tilde{\Gamma}^f}(\ve-\tilde{\ve}_d)\left[\dot{\tilde{\Gamma}}^f\left(\dot{\ve}_d+(\ve-\tilde{\ve}_d)\frac{\dot{\Gamma}}{\Gamma}\!\right)\right.\right.\nonumber\\
&&\left.+\dot{\tilde{\ve}}_d\dot{\Gamma}\calb^{f^2}\Bigg]-\frac{1}{2}\left(\frac{\dot{\Gamma}\,\dot{\tilde{\Gamma}}^f}{\Gamma}-\sum_\alpha\frac{\dot{\Gamma}_\alpha\dot{\tilde{\Gamma}}_\alpha^f}{\Gamma_\alpha}\right)\!\!\right\}\nonumber\\
&&+\frac{T}{2\pi\hbar^2\omega}\!\int^\tau_0\!\!dt\Big\{{\dot{\ve}_d}\left[C_2^\omega-\Delta{\calb^2}^{\omega}\right]\\
&&\left.+\frac{\dot{\Gamma}\calb^{f^2}}{\Gamma}\left[C_1^\omega-\Delta\lambda^{\omega}-\lambda^f\frac{\Delta{\calb^2}^{\omega}}{\calb^{f^2}}\right]\right\}.\nonumber
\ea
The expressions for the vectors $\vec{C}^{V,\omega}(t)$ and the corrections $\Delta{\lambda}^{V,\omega}(t)$ and $\Delta{\calb^2}^{V,\omega}(t)$ can be found, respectively, in Eqs. (\ref{c}), (\ref{correct1}) and (\ref{correct2}). 
In the above two equations we have avoided the explicit energy and temporal dependences of the integrands in order to make the expressions more compact. 

Now, in order to prove the validity of the Onsager reciprocal relation in Eq. (\ref{onsager}), we will show that  $\Delta=L_{12}+L_{21}=0$. For that, we start from Eqs. (\ref{c}), (\ref{l12}) and (\ref{l21}), and find
\ba\label{Delta}
\Delta &=&\frac{T}{eV h}\int^{\tau}_{0}\!\!dt\Bigg\{\left(C_2^V(t)\dot{\lambda}^f(t)+C_1^V(t)\dot{\calb}^{f^2}(t)\right)\nonumber\\
&&+{\dot{\ve}_d(t)}\left[C_2^V(t)-\Delta{\calb^2}^{V}\!\!(t)\right]\\
&&\left.+\frac{\dot{\Gamma}(t)\calb^{f^2}\!(t)}{\Gamma(t)}\left[C_1^V(t)-\Delta\lambda^{V}\!(t)-\lambda^f(t)\frac{\Delta{\calb^2}^{V}\!\!(t)}{\calb^{f^2}\!(t)}\right]\right\}.\nonumber
\ea
Then, by performing the temporal derivative of the system of non-linear equations in Eq. (\ref{setfrozen}), we get another set of linear equations for finding $\dot{\lambda}^f(t)$ and $\dot{\calb^{f^2}}(t)$ in terms of the {\it{frozen}} ${\lambda}^f(t)$ and ${\calb^{f^2}}(t)$ and the derivatives in time of the applied {\it{ac}}-fields, $\dot{\ve}_d(t)$ and $\dot{\Gamma}(t)$. The solutions read
\ba\label{derivlambda}
\dot{\lambda}^f(t)&=&{\dot{\Gamma}(t)}\frac{(M_{22}(t)\lambda^f(t)-M_{12}(t)\calb^{f^2}(t))}{\Gamma(t)\mbox{det}[\hat{M}(t)]}\nonumber\\
&&+\dot{\ve}_d(t)\left(\frac{M_{22}(t)}{\mbox{det}[\hat{M}(t)]}-1\right),
\ea
and
\ba\label{derivb}
\dot{\calb^{f^2}}(t)&=&-\frac{\dot{\Gamma}(t)}{\Gamma(t)}\calb^{f^2}\!(t)\!\!\left(1\!+\!\frac{\lambda^f(t)}{\calb^{f^2}(t)}\frac{M_{21}(t)}{\mbox{det}[\hat{M}(t)]}\!-\!\frac{M_{11}(t)}{\mbox{det}[\hat{M}(t)]}\right)\nonumber\\
&&-\dot{\ve}_d(t)\frac{M_{21}(t)}{\mbox{det}[\hat{M}(t)]},
\ea
where the elements of the matrix $\hat{M}(t)$ are the same as in Eq. (\ref{m}). Finally, by replacing Eqs. (\ref{derivlambda}) and (\ref{derivb}) into (\ref{Delta}), and using the expressions for the corrections in (\ref{correct1}) and (\ref{correct2}), we find that at every instant of time
\ba
&&C_2^V\!(t)\dot{\lambda}^f(t)\!+\!C_1^V\!(t)\dot{\calb}^{f^2}\!(t)=\!-{\dot{\ve}_d(t)}\!\left(\!C_2^V(t)-\Delta{\calb^2}^{V}\!\!(t)\right)\nonumber\\
&&-\frac{\dot{\Gamma}(t)\calb^{f^2}\!(t)}{\Gamma(t)}\left(C_1^V(t)-\Delta\lambda^{V}\!(t)-\lambda^f(t)\frac{\Delta{\calb^2}^{V}\!\!(t)}{\calb^{f^2}\!(t)}\right),
\ea
and therefore $\Delta=0$. 

Analogously, by replacing Eqs. (\ref{derivlambda}) and (\ref{derivb}) into (\ref{l22sup}), we get Eq. (\ref{l22}).

\end{document}